\documentclass[pre,superscriptaddress,twocolumn,showpacs]{revtex4}\def\tc{yy}
\usepackage{psfig}
\begin{document}
\def\<{\left\langle}
\def\>{\right\rangle}
\def\({\left (}
\def\){\right )}
\def\[{\left [}
\def\]{\right ]}
\def\i{{\rm i}}
\def\e{{\rm e}}
\def\/{\over}
\def\figdir{.}
\title{Effective Coupling for Open Billiards}
\author{Konstantin Pichugin}
\affiliation{Institute of Physics, Czech Academy of Sciences, 
Cukrovarnicka 10, Prague, Czech Republic}
\author{Holger Schanz}
\email{holger@chaos.gwdg.de}
\affiliation{Max-Planck-Institut f\"ur Str\"omungsforschung und Institut
f{\"u}r Nichtlineare Dynamik\\ der Universit{\"a}t G{\"o}ttingen, 
Bunsenstra{\ss}e 10, D-37073 G\"ottingen, Germany}
\author{Petr \v{S}eba}
\affiliation{Institute of Physics, Czech Academy of Sciences, 
Cukrovarnicka 10, Prague, Czech Republic}
\affiliation{Department of Physics, Pedagogical University, 
Hradec Kralove, Czech Republic}
\date{April 9, 2001}
\pacs{03.65.N, 05.45.Mt,72.20.D}
\begin{abstract}
We derive an explicit expression for the coupling constants of individual
eigenstates of a closed billiard which is opened by attaching a waveguide. The
Wigner time delay and the resonance positions resulting from the coupling
constants are compared to an exact numerical calculation. Deviations can be
attributed to evanescent modes in the waveguide and to the finite number of
eigenstates taken into account. The influence of the shape of the billiard and
of the boundary conditions at the mouth of the waveguide are also
discussed. Finally we show that the mean value of the dimensionless coupling
constants tends to the critical value when the eigenstates of the billiard
follow random-matrix theory.
\end{abstract}
\maketitle
\section{Introduction}
During the last years quantum chaotic scattering was a field of intense
research. A great deal of the results obtained was based on the projection
operator formalism due to Feshbach, Weidenm{\"u}ller and others
\cite{MW69,Feshbach,Dit00}. In this approach the scattering system is
decomposed into a closed subsystem described by the internal Hamiltonian
$H_{\rm in}$ with discrete bound states $n=1\dots N$ and a continuum of
external scattering states labeled by the energy $E$ and an index
$\lambda=1\dots\Lambda(E)$ corresponding to different open scattering
channels. The coupling between the internal and external subsystems is then
incorporated by an operator with matrix elements $W_{n,\lambda}(E)$. The
S-matrix of the complete system can be expressed in terms of these matrix
elements and the Hamiltonian $H_{\rm in}$.  This relation can be cast into the
form
\begin{equation}\label{sk} S={I-\i\,K\over I+\i\,K}\end{equation}
\begin{equation}\label{sk2}K=\pi W^{\dag}{I\over E-H_{\rm in}}W\,.
\end{equation}
Here, $S$ and $K$ are energy-dependent square matrices of dimension
$\Lambda\times\Lambda$, and $W$ has dimension $N\times\Lambda$.  While this
setting is very general, the tools developed for the subsequent analysis of
the properties of the S-matrix require additional assumptions. In particular,
the energy thresholds for the opening of new scattering channels are usually
neglected. As a consequence the energy dependence of the coupling matrix
$W(E)$ can be considered weak and then the S-matrix (\ref{sk}) can be
rewritten in terms of an $N\times N$ effective non hermitian Hamiltonian
$H_{\rm eff}$
\begin{equation}\label{seffh}
S=I-2\pi\i W^{\dag}{I\over E-H_{\rm eff}}W 
\end{equation}
\begin{equation}\label{seffh2}
H_{\rm eff}=H_{\rm in}-\i\pi\,W\,W^{\dag}\,.
\end{equation}
This canonical formalism \cite{MW69,Feshbach,Dit00} for expressing the
S-matrix is sometimes denoted as the Heidelberg description of scattering, and
we use this name to distinguish it from an S-Matrix obtained directly, i.~e.\
without reference to any auxiliary closed system.

When the internal Hamiltonian in (\ref{seffh2}) describes a chaotic system, it
is justified to replace it by a random matrix \cite{BGS84}, and by performing
an average over the appropriate ensemble a statistical theory for the S-matrix
is obtained which allows to calculate quantities of interest such as
correlation functions or the distribution of Wigner delay times and resonance
poles \cite{MPS85,LW91,FS97,GMW98}. It was found that the results of such an
approach are to a large extent independent of the detailed structure of the
matrix $W$, but they do depend on the dimensionless mean coupling strength
\begin{equation}\label{coupling}
g=\pi^{2}\,{\langle|W_{n,\lambda}|^{2}\rangle_{n,\lambda}\over D}\,.
\end{equation}
Here, $D$ is the mean energy level spacing of the internal subsystem, and the
average $\langle\dots\rangle_{n,\lambda}$ is taken over the internal states
$n$ and all open scattering channels $\lambda$. 

For example, when the coupling constant (\ref{coupling}) exceeds the critical
value $g=1$ and the number of scattering channels is small compared to the
total number of states, a counterintuitive shrinking of the widths of most
resonances with increasing coupling is observed \cite{SZ89,SZ92,Rot91}. For
each attached scattering channel only one of the resonance widths grows
further with the coupling $g$.  The resulting redistribution of S-matrix poles
was coined resonance trapping.

Chaotic billiards with attached wave\-guides are con\-sidered as paradigm for
chaotic scat\-ter\-ing \cite{DS92a}. They are relevant as theoretical models
for understanding the transport properties of mesoscopic semiconductor
structures \cite{M+92} or experimental results on microwave scattering in flat
resonators \cite{SS90,P+00}. Also signatures of resonance trapping were recently
observed in billiards both, numerically \cite{P+98,R+00,S+00} and in microwave
resonator experiments \cite{P+00}. 

However, to our knowledge there is no theory which maps a given billiard to an
effective Hamiltonian with overcritical coupling, thus really establishing a
connection between the numerically and experimentally observed phenomena and
the results on resonance trapping obtained within the formalism
(\ref{sk})-(\ref{coupling}).

Motivated by this situation, it is the purpose of the present paper to discuss
the application of the Heidelberg approach to open billiards in some detail
and to answer questions such as: How can (\ref{sk}), (\ref{seffh}) be derived
for a billiard, what kind of approximations are involved and what is the
resulting expression for the coupling constants $W_{n,\lambda}(E)$? What is
the influence of the choice of the internal subsystem which is {\em not
unique} for a given scattering system?

Using the expression for the coupling constants $W_{n,\lambda}$ to be derived
in section II we will then address the effective coupling constant for a
typical chaotic billiard with an attached waveguide. We show in section III
that {\em in the semiclassical regime}, and when no tunneling barriers
obstruct the waveguides, the coupling strength is fixed at the critical value
$g=1$, independently of the size or the precise geometry of the billiard and
of its openings. A numerical verification of our results is contained in
section IV, followed by a short discussion of the implications of our
findings.

\section{Coupling Constants for Individual Levels}
\def\figSketch{
 \begin{figure}[htb]
  \centerline{\psfig{figure=\figdir/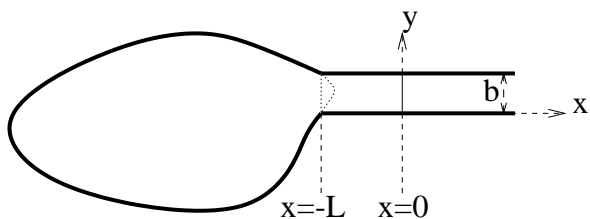,width=\fw}}
  \caption{\label{sketch} A scattering system consisting of an infinite
  waveguide and a cavity is shown with bold lines. Various possibilities to add
  a wall and obtain a closed billiard are shown with thin lines (solid and
  dotted).}
 \end{figure}
}
\if\tc{\def\fw{78mm}\figSketch}\fi
We consider a situation as shown in Fig.~\ref{sketch}. A scattering system is
formed in two dimensions by an infinite waveguide of width $b$ and and an
arbitrary cavity. Inside the system the potential is identically 0. We set
$\hbar=2m=1$ and $E=k^{2}$ such that the stationary Schr{\"o}dinger equation
reduces to the Helmholtz equation 
\begin{equation}\label{Helmholtz}
(\Delta+k^{2})\Psi(x,y)=0\,,
\end{equation}
with $\Delta=\partial^{2}/\partial_{x}^{2}+\partial^{2}/\partial_{y}^{2}$. On
the boundary of the scattering system (bold solid line in Fig.~\ref{sketch})
we require Dirichlet b.~c. $\Psi=0$. This boundary condition and also the
precise geometry of the system are by no means essential, the following
generalizes e.~g.\ immediately to a cavity with more than one attached lead or
Neumann b.~c. $\partial/\partial_{n}\Psi=0$.  In Fig.~\ref{sketch} we have
shown several possibilities to define a closed billiard which corresponds to
the scattering system in question (solid and dotted thin lines). We will
restrict the discussion to the case shown with a solid line: We require that
the boundary of the internal system is located inside the attached waveguide
and that it consists of a transversal straight line on which either Dirichlet
or Neumann boundary conditions are imposed to close the system. Clearly, even
this restriction makes the correspondence between the scattering system and
the auxiliary internal system not unique because the exact position of the
closure along the waveguide is variable. We use coordinates where this closure
is at $x=0$ while the matching between waveguide and cavity is at $x=-L$
$(L>0)$.

In the region of the attached waveguide ($x\ge -L$) we can decompose any
function into transversal modes \begin{equation}\label{transmode}
\phi_{\lambda}(y)=\sqrt{2\/b}\sin\(\lambda{\pi\,y\/b}\)
\qquad(\lambda=1,2,\dots)\,,
\end{equation}
because these functions form a complete and orthonormal basis on the interval
$(0,b)$ according to
\begin{equation}\label{complete}
\sum_{\lambda=1}^{\infty}\phi_{\lambda}(y)\phi_{\lambda}(y')=\delta(y-y')
\qquad (0<y,y'<b)
\end{equation}
and
\begin{equation}\label{orthonormal}
\int_{0}^{b}{\rm d}y\,\phi_{\lambda}(y)\phi_{\lambda'}(y)=\delta_{\lambda\lambda'}\,.
\end{equation}
The most general solution of the Helmholtz
equation is a superposition of the scattering states $\Psi_{\lambda}(x,y)$ which
consist of a single incoming wave in transversal mode $\lambda$ and the
corresponding outgoing modes given by the S-matrix of the system
\begin{equation}\label{sstate} \Psi_{\lambda}(x,y)=
\phi_{\lambda}(y){\e^{-\i k_{\lambda}x}\over \sqrt{k_{\lambda}}}
+\sum_{\lambda'}S_{\lambda'\lambda}\,
\phi_{\lambda'}(y)\,{\e^{+\i k_{\lambda'}x}\over \sqrt{k_{\lambda'}}}\,.
\end{equation}
The longitudinal wave number $k_{\lambda}=\sqrt{k^{2}-(\lambda\pi/b)^{2}}$ is
real for $\lambda\le\Lambda=[kb/\pi]$, where $[\dots]$ denotes the integer
part. These modes are called open or travelling, and the
$\Lambda\times\Lambda$ matrix $S_{\lambda'\lambda}$ corresponding to the open
modes is the unitary S-matrix we are interested in. For $\lambda>\Lambda$ the
momentum along the waveguide is imaginary. These modes are called closed or
evanescent. In the scattering state $\Psi_{\lambda}$ with $\lambda\le\Lambda$
the evanescent outgoing modes describe exponentially decaying contributions
which modify the wave function in the vicinity of the mouth of the
waveguide. Evanescent incoming modes are exponentially increasing into the
waveguide and thus unphysical for the scattering system.

When the evanescent modes are included, the S-matrix becomes an
infinite-dimensional operator which is no longer unitary. It is possible to
construct an eigenstate of the closed billiard by a superposition of the
scattering states (\ref{sstate}) including evanescent modes: Suppose $S(E)$
has an eigenvalue unity at some energy $E_{n}=k_{n}^{2}$ and let
$a_{n,\lambda}$ be the components of the corresponding eigenvector. Then the
linear combination of scattering states
\begin{eqnarray}\label{billefN}
\Psi_{n}^{\rm (N)}(x,y)&=&\sum_{\lambda}{a_{n,\lambda}^{\rm (N)}\/2}\Psi_{\lambda}(x,y)
\nonumber\\ &=&
\sum_{\lambda}{a^{\rm (N)}_{n,\lambda}\/\sqrt{k_{n,\lambda}}}\phi_{\lambda}(y)\cos(k_{n,\lambda}\,x)
\end{eqnarray}
$(-L\le x\le 0)$ satisfies Neumann b.~c. at $x=0$ and it is thus indeed an
eigenfunction of the billiard because it satisfies (\ref{Helmholtz}) and the
remaining boundary conditions by construction. The normalization of the
S-matrix eigenvector $a_{n,\lambda}$ in (\ref{billefN}) is {\em not} unity but
rather determined by the normalization of the billiard eigenfunction
$\Psi_{n}^{\rm (N)}(x,y)$.  If Dirichlet boundary conditions are required at
$x=0$ the same argument can be repeated for an eigenvalue $-1$ of the S-matrix
and we have \begin{eqnarray}\label{billefD} \Psi_{n}^{\rm (D)}(x,y)
&=&\sum_{\lambda}{a_{n,\lambda}^{\rm (D)}\/2\i}\Psi_{\lambda}(x,y) \nonumber\\
&=& \sum_{\lambda}{a^{\rm
(D)}_{n,\lambda}\/\sqrt{k_{n,\lambda}}}\phi_{\lambda}(y)\sin(k_{n,\lambda}\,x)
\end{eqnarray}
$(-L\le x\le 0)$. Consequently the spectrum of the billiard closed with
Neumann or Dirichlet b.~c.\ can be found from the secular equation
\begin{equation}\label{seceq}
\det(I\mp S(E))=0\,,
\end{equation}
which was first derived by Doron and Smilansky \cite{DS92b}. In a sense we
will in the following invert this so-called scattering approach to the
quantization of billiards: we will express the S-matrix in terms of the
eigenvalues and eigenfunctions of the closed system.

For this purpose consider the Green function of the closed billiard which is
defined as the resolvent of $-\Delta$ in the space of functions which satisfy
the boundary conditions of the billiard. In position representation this
definition can be expressed by the inhomogeneous Helmholtz equation
\begin{equation}\label{ihHelmholtz} 
(\Delta+k^{2})\,G({\bf r};{\bf r'},k)=\delta({\bf r}-{\bf r'})\,.
\end{equation}
In the eigenbasis of the billiard the Green function reads
\begin{equation}\label{GFdeco}
G({\bf r};{\bf r'},k)=
\sum_{n=1}^{\infty} {\Psi_{n}^{*}({\bf r'})\Psi_{n}({\bf r})\over
k^{2}-k_{n}^{2}}\,,
\end{equation}
which can be verified using $\Delta\Psi_{n}({\bf r})=-k_{n}^{2}\Psi_{n}({\bf
r})$ and the completeness of the functions $\Psi_{n}$ inside the billiard. For
$\bf r$ and $\bf r'$ inside the waveguide we can expand the Green function
with respect to the transversal modes $\phi_{\lambda}(y)$ and find as the
general form of a solution of (\ref{ihHelmholtz})
\begin{eqnarray}\label{GFdeco1}
G({\bf r},{\bf r'},k)&=&{1\/2\i}\sum_{\lambda\lambda'}
{\phi_{\lambda }(y )\/\sqrt{k_{\lambda }}}
{\phi_{\lambda'}(y')\/\sqrt{k_{\lambda'}}}
\\&&\hspace*{-10mm}\times
\Bigg[\delta_{\lambda\lambda'} \e^{\i k_{\lambda}\left|x-x'\right|}+
 \sum_{s,s'=\pm}G_{\lambda\lambda'}^{ss'}(k) \e^{\i sk_{\lambda}x+\i s'k_{\lambda'}x'}
\Bigg]\,.
\nonumber
\end{eqnarray}
Indeed, the first term inside the brackets gives rise to one particular
solution of the inhomogeneous equation (\ref{ihHelmholtz}), while the second
term with the unknown matrices $G^{++}$, $G^{+-}$, $G^{-+}$ and $G^{--}$
represents the most general solution of the corresponding homogeneous
Helmholtz equation (\ref{Helmholtz}). The unknown coefficients must be
determined such that the Green function satisfies also the boundary conditions
inside the cavity and on the transversal closure of the waveguide. For this
purpose assume first $x\ge x'$ and consider $\bf r$ as a fixed parameter. Then
$G({\bf r};{\bf r'},k)$ as a function of $\bf r'$ should satisfy the
homogeneous Helmholtz equation with the boundary conditions of the scattering
system, i.~e.\ it can be written as a superposition of the scattering states
$\Psi_{\lambda}({\bf r'})$ defined in Eq.~(\ref{sstate}). On the other hand, when
$\bf r'$ is fixed, the Green function as a function of $\bf r$ satisfies the
boundary conditions (Neumann or Dirichlet) at $x=0$ where the closed billiard
is separated from the waveguide by the additional straight wall. Thus, it must
be a superposition of the functions
\begin{equation}\label{PsiND} \Psi_{\lambda}^{\rm N/D}({\bf
r})= {\phi_{\lambda}(y)\over \sqrt{k_{\lambda}}}
(\e^{\i k_{\lambda}x}\pm \e^{-\i k_{\lambda}x})\,,
\end{equation}
which are in fact the scattering states for a semi-infinite waveguide with
Neumann or Dirichlet b.c. at one end.  Consequently, the Green function has
the form \begin{equation}\label{GFdeco2} G({\bf r};{\bf r'},k)={1\over
2\i}\sum_{\lambda,\lambda'} \Psi_{\lambda}^{\rm N/D}({\bf
r})\,g_{\lambda\lambda'}(k)\,\Psi_{\lambda'}({\bf r'})\,,
\end{equation}
with another set of undetermined coefficients $g_{\lambda\lambda'}$. Expanding
(\ref{GFdeco2}) into transversal modes and comparing to (\ref{GFdeco1}) we
obtain
\begin{equation}\label{relations}
\begin{array}{ll}
G^{++}=g\,S & G^{+-}=g-I\cr
G^{-+}=\pm g\,S & G^{--}=\pm g
\end{array}\,.
\end{equation}
We can now repeat this argumentation under the opposite assumption $x<x'$ and
find again the relations (\ref{relations}) but with $G^{+-}$ and $G^{-+}$
exchanged. This can be regarded as a consequence of the symmetry of the Green
function with respect to its two arguments which in turn follows from
time-reversal symmetry. We conclude $G^{+-}=G^{-+}=g-I=\pm g\,S$ and hence
\begin{equation}\label{gS} g(k)=(I\mp S(k))^{-1}\,.
\end{equation}
Note that $g(k)$ and thus the Green function diverges as expected at the
solutions of the secular equation (\ref{seceq}), i.~e., when $k$ corresponds
to an eigenvalue of the closed billiard.

Using Eq.~(\ref{gS}) we can now directly relate the S-matrix to the
transversal expansion coefficients of the Green function at the closure of the
billiard. We define for $x,x'$ inside the waveguide
\begin{equation}
G_{\lambda,\lambda'}(x,x')=
\int_{0}^{b}{\rm d}y\,{\rm d}y'
\phi_{\lambda}(y)\,G(x,y;x',y';k)\,\phi_{\lambda'}(y')
\end{equation}
and
\begin{equation}\label{KN}
K_{\lambda\lambda'}^{\rm (N)}=\sqrt{k_{\lambda}k_{\lambda'}}\,G_{\lambda,\lambda'}(0,0)
\end{equation}
\begin{equation}\label{KD}
K_{\lambda\lambda'}^{\rm (D)}=
{1\/\sqrt{k_{\lambda}k_{\lambda'}}}\,
\left.{\partial^{2}\over
\partial_{x}\partial_{x'}}G_{\lambda,\lambda'}(x,x')\right|_{x=x'=0}
\end{equation}
and derive from (\ref{GFdeco2}) using (\ref{gS})
\begin{equation}\label{KSN}
\i\,K^{\rm (N)}={I+S\over I-S}\qquad(\mbox{Neumann b.~c. at }x=0)
\end{equation}
\begin{equation}\label{KSD}
\i\,K^{\rm (D)}={I-S\over I+S}\qquad(\mbox{Dirichlet b.~c. at }x=0)\,.
\end{equation}
Obviously, $K^{\rm (N)}=0$ for Dirichlet b.~c.\ and $K^{\rm (D)}=0$ for Neumann
b.~c.\ at $x=0$. Eqs.~(\ref{KSN}), (\ref{KSD}) can be inverted and yield
\begin{equation}\label{SKND}
S
=-{I-\i\,K^{\rm (N)}\over I+\i\,K^{\rm (N)}}
=+{I-\i\,K^{\rm (D)}\over I+\i\,K^{\rm (D)}}\,,
\end{equation}
which is now in the form of (\ref{sk}) (for Neumann b.~c.\ up to an irrelevant
constant phase). We can now proceed to determine the corresponding coupling
constants $W^{\rm (N/D)}_{n,\lambda}$ by representing the K-matrix in the
eigenbasis of the billiard. From (\ref{sk2}) we have
\begin{equation}\label{KW}
K_{\lambda\lambda'}=\pi\,\sum_{n=1}^{\infty}\,
{W_{n,\lambda}^{*}\,W_{n,\lambda'}\/k^{2}-k_{n}^{2}}
\end{equation}
and from (\ref{GFdeco}), (\ref{KN}), (\ref{KD}) we find
\begin{equation}\label{KPsiN}
K_{\lambda\lambda'}^{\rm (N)}=\sqrt{k_{\lambda}k_{\lambda'}}
\sum_{n=1}^{\infty}\,{
\Psi^{\rm (N)*}_{n,\lambda}(0)
\Psi^{\rm (N)}_{n,\lambda'}(0)
\/k^{2}-k_{n}^{2}}
\end{equation}
\begin{equation}\label{KPsiD}
K_{\lambda\lambda'}^{\rm (D)}={1\/\sqrt{k_{\lambda}k_{\lambda'}}}
\sum_{n=1}^{\infty}\,{
{\partial\/\partial x}\Psi^{\rm (D)*}_{n,\lambda}(0)
{\partial\/\partial x}\Psi^{\rm (D)}_{n,\lambda'}(0)
\/k^{2}-k_{n}^{2}}
\end{equation}
where we have introduced the projections
\begin{equation}\label{billefpr}
\Psi^{\rm (N/D)}_{n,\lambda}(x)=\int_{0}^{b}{\rm d}y\,\phi_{\lambda}(y)\,
\Psi^{\rm (N/D)}_{n}(x,y)
\end{equation}
of the eigenfunctions of the closed billiard onto the transversal modes of the
waveguide. The values of the coupling constants follow from comparing
(\ref{KW}) to (\ref{KPsiN}) and (\ref{KPsiD})
\begin{equation}\label{WN}
W^{\rm (N)}_{n,\lambda}=\sqrt{k_{\lambda}\/\pi}\,\Psi^{\rm (N)}_{n,\lambda}(0)\,,
\end{equation}
\begin{equation}\label{WD}
W^{\rm (D)}_{n,\lambda}={1\/\sqrt{k_{\lambda}\pi}}\,
{\partial\/\partial x}\Psi^{\rm (D)}_{n,\lambda}(0)\,.
\end{equation}
This form of the dependence of the coupling constants on the internal wave
functions is not surprising: also within perturbation theory the coupling
depends on the value of the wave function at the point where the system is
opened or on its normal derivative for Neumann and Dirichlet boundary
conditions, respectively. However, in the situation we consider
perturbation theory is not applicable and, in particular, the precise value of
the prefactor in the coupling constants (\ref{WN}), (\ref{WD}) could only be
obtained from the derivation given in this section.

The representation (\ref{SKND}) of the S-matrix in terms of the K-matrices
(\ref{KPsiN}), (\ref{KPsiD}) is exact, when all transversal modes are
included. However, usually one is interested only in the
$\Lambda\times\Lambda$ unitary part of the S-matrix, and this is only {\em
approximately} given by (\ref{SKND}) when the K-matrix is restricted to open
modes. In \cite{DS92b,SS95} the effect of this so-called semiquantal
approximation for the accuracy of eigenvalues within the scattering approach
to quantization was investigated numerically. It becomes negligible when the
energy $E$ is sufficiently far from the threshold for the opening of a new
channel. Under this restriction we can consider $W^{\rm (N/D)}_{n,\lambda}$ as
the coupling constants corresponding to the unitary part of the S-matrix. When
$E$ approaches a threshold, Eq.~(\ref{seffh}) breaks down, since the
energy-dependence of the coupling constants can no longer be neglected. We
will not consider this case here.

\section{The Mean Coupling Strength}
Given the explicit values (\ref{WN}), (\ref{WD}) for the coupling constants
between individual states and individual scattering channels we are now going
to derive an estimate for the dimensionless coupling strength (\ref{coupling})
in the semiclassical limit and neglecting evanescent modes. Since the concept
of a mean coupling strength is not well-defined for infinitely many internal
states, the internal Hamiltonian $H_{\rm in}$ entering (\ref{sk2}),
(\ref{seffh2}) should for this purpose be cut to some finite matrix including
only states which are close enough in energy $k_{n}\sim k$. In particular this
means that we can replace the momenta along the waveguide $k_{n,\lambda}$ in
(\ref{billefN}), (\ref{billefD}) by their on-shell values $k_{\lambda}$. The
resulting approximate expansions of the billiard eigenfunctions are projected
onto the transversal modes according to (\ref{billefpr}) and inserted into
(\ref{WN}), (\ref{WD}) which simplify to
\begin{equation}\label{WND} W^{\rm
(N/D)}_{n,\lambda}={a^{\rm (N/D)}_{n,\lambda}\/\sqrt{\pi}}\,.
\end{equation}
At this point it is necessary to determine the average magnitude of the
coefficients $a^{\rm (N/D)}_{n,\lambda}$. We assume that the classical
dynamics of the billiard is chaotic. In the semiclassical limit this means
that the quantum ergodicity theorem applies to the eigenstates of the
billiard, i.~e.\ in particular the probability density integrated over an
arbitrary region of the billiard tends to the relative area of that
region. Applied to the part of the billiard inside the waveguide we find
\begin{eqnarray}\label{normalization} {bL\over A}&\approx&\int_{0}^{b}{\rm
d}y\,\int_{-L}^{0}{\rm d}x\, |\Psi^{\rm (N)}_{n}(x,y)|^{2}
\nonumber\\&=&\int_{-L}^{0}{\rm d}x\, \sum_{\lambda}{|a_{n,\lambda}|^{2}\over
k_{\lambda}}\cos^{2}(k_{\lambda}x) \nonumber\\ &\approx&
\<|a|^2\>{L\/2}\int_{0}^{kb/\pi}{\rm d}\lambda\, \sqrt{1\/k^2-(\lambda\pi/b)^2}
\nonumber\\&=& \<|a|^2\>{bL\/4}\,,
\end{eqnarray}
where $A$ denotes the total area of billiard. In the second line we have
inserted the normal mode decomposition into $|\Psi^{\rm (N)}_{n}(x,y)|^{2}$.
The orthonormalization of the transversal modes was then used to restrict the
resulting double sum over modes to diagonal terms.  In the third
line $\<\cos^2\>=1/2$ was used ($\<\sin^2\>=1/2$ in the completely analogous
calculation for Dirichlet b.c.), and the sum over modes was approximated by a
continuous integral. This is justified when the number of modes is large,
i.~e.\ in the semiclassical limit. Using the resulting constraint on the 
normalization of the coefficients of the billiard eigenfunctions in the
transversal basis implied by (\ref{normalization}), 
$\<|a^{\rm (N/D)}_{n,\lambda}|^2\>={4/A}$, we find 
\begin{equation}
\<|W^{\rm (N/D)}_{n,\lambda}|^2\>={4\/A\pi}
\end{equation}
According to (\ref{coupling}) the average coupling between the internal states
and the continuum must be normalized by the mean level spacing $D$ of the
billiard which is the only independent energy scale of the system. To leading
semiclassical order we have Weyl's law $D=4\pi/A$ \cite{BH76} which finally
results in \begin{equation}\label{g1} g=1\,.
\end{equation}
\section{Numerical Results}
\def\figGeom{
 \begin{figure}[htb]
  \centerline{\psfig{figure=\figdir/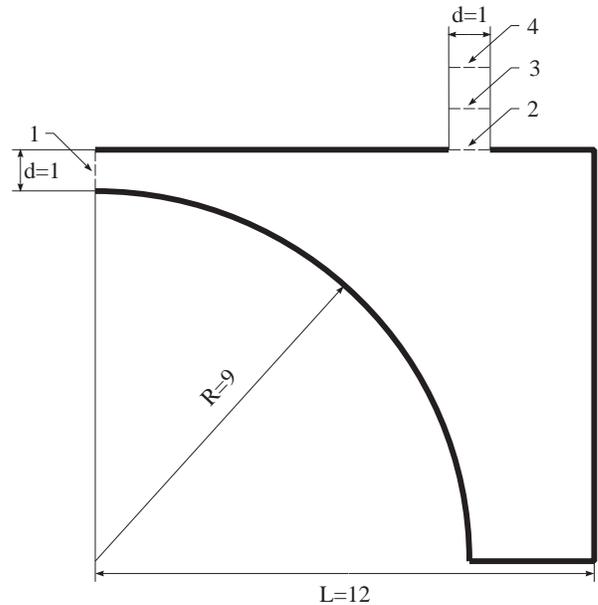,width=\fw}}
  \caption{\label{geom}Geometry of the billiard. Four different possibilities
  to attach the waveguide are used.}
 \end{figure}
}
\if\tc{\def\fw{78mm}\figGeom}\fi
To check the validity of the Heidelberg approach for a quantum billiard we
have performed direct numerical calculations for a Sinai billiard connected to
a single waveguide (see Fig.~\ref{geom}). First, we evaluated numerically the
$1300$ lowest eigenvalues and eigenvectors of the closed system with Dirichlet
and Neumann b.c.\ at the boundary segment to which the waveguide was attached.
Using the expressions (\ref{billefpr})-(\ref{WD}) we calculated the elements of
the coupling matrix $W^{\rm(N/D)}_{n,\lambda}$. The first observation is that
the distribution of the numerically obtained values of
$W^{\rm(N/D)}_{n,\lambda}$ is fairly close to a Poisson distribution, that is
expected for a fully random coupling. Hence the elements of the coupling
matrix $W$ evaluated with help of the formulae (\ref{billefpr})-(\ref{WD})
resemble the random coupling standardly used in the Heidelberg approach.

Knowing $W$ and using Eqs.~(\ref{sk}), (\ref{sk2}) we evaluated---as the next
step---the $S$-matrix and compared it to the $S$-matrix obtained by a direct
method based on the numerical solution of the underlying Schr{\"o}dinger
equation (see Ref.~\cite{A91} for details). In order to visualize the
differences between these two $S$-matrices we compared first the corresponding
Wigner-Smith time delays
\begin{equation}\label{WSTD} \tau(E)=\frac{\i}{M} {\rm Tr}\left(
\frac{\partial {S^\dag}(E)}{\partial E} {S}(E)\right)\,,
\end{equation}
where $M$ is the number of the open channels inside the
waveguide. The results are plotted in Fig.~\ref{TD_low}.

\def\figTDlow{
 \begin{figure}[htb]
  \centerline{\psfig{figure=\figdir/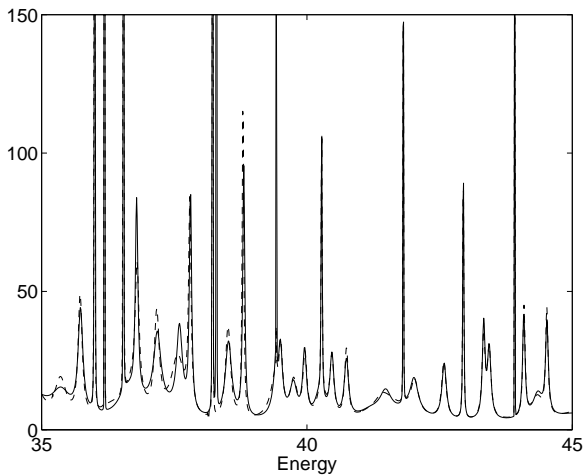,width=\fw}}
  \caption{\label{TD_low}The Wigner-Smith time delay obtained for the
  Heidelberg $S$-matrix with Neumann b.c.\ (dashed line) compared with the
  result of a direct evaluation (solid line).  In the displayed case the
  waveguide was attached to the boundary No. $1$.  }
 \end{figure}
}
\if\tc{\def\fw{78mm}\figTDlow}\fi
The figure shows an outstanding  agreement of the two approaches for low
energies. The Heidelberg approach describes the positions as well as the widths
of the narrow  resonances with high accuracy. The agreement is good even close
to the threshold energies of the individual channels. For higher energies,
however, the difference between the two time delay functions increases due to
the limited number of internal states included into the evaluation of the
Heidelberg $S$-matrix. Similar results (not displayed) were obtained also with
the waveguide attached to the boundaries No. $2$, $3$ and $4$ respectively.  In
all these cases the Neumann matching procedure was used.

On the other hand, for Dirichlet matching (Eq.~\ref{WD}) the results change
drastically. In this case the agreement is not good even for low
energies. This may seem surprising because the derivations of the previous
sections were entirely parallel for Neumann and Dirichlet b.~c. However, an
important difference is hidden in the convergence properties of the spectral
decompositions of the K-matrix (\ref{KPsiN}), (\ref{KPsiD}) as we shall
explain now. Projecting (\ref{billefN}), (\ref{billefD}) onto transversal
mode $\lambda$  we find
\begin{equation}
\Psi^{\rm (N)}_{n,\lambda}(0)={a^{\rm (N)}_{n,\lambda}\/\sqrt{k_{n,\lambda}}}
\qquad
{\partial\over \partial x}\Psi^{\rm (D)}_{n,\lambda}(0)=\sqrt{k_{n,\lambda}}\,
a^{\rm (D)}_{n,\lambda}\,.
\end{equation}
Since the coefficients $a^{\rm (N/D)}_{n,\lambda}$ are according to
(\ref{normalization}) of order $2/\sqrt{A}$ which is independent of $n$, we
have from (\ref{WN}), (\ref{WD})  
\begin{equation}
W^{\rm (N)}_{n,\lambda}\sim\sqrt{k_{\lambda}\/k_{n,\lambda}}\qquad
W^{\rm (D)}_{n,\lambda}\sim\sqrt{k_{n,\lambda}\/k_{\lambda}}\,.
\end{equation}
For a 2D billiard $k_{n}, k_{n,\lambda}=O(\sqrt{n})$ $(n\to\infty)$ such that
the terms in the infinite spectral sum (\ref{KW}) decay asymptotically as
$n^{-3/2}$ for Neumann and as $n^{-1/2}$ for Dirichlet b.c. Hence the
convergence is absolute for Neumann b.c. while (\ref{KPsiD}) converges only
conditionally. As a consequence, the numerically necessary cut-off in the
summation over the internal states $n$ introduces large errors for Dirichlet
b.c. 

For the following considerations we will concentrate on Neumann b.c.  As
already mentioned, the choice of the internal and external part of the system
is not unique, since an arbitrary part of the ideal waveguide can be
considered part of the internal system.  Increasing the length of the
waveguide included, we decrease in fact the influence of the evanescent modes
since they are exponentially vanishing inside waveguide. We have checked this
relation and evaluated the time delay functions also for various waveguide
parts included into the internal system. The results remain practically
unchanged regardless on the length of the included part.  This demonstrates
the small influence of the evanescent modes on the resulting Heidelberg
$S$-matrix.

Knowing the coupling matrix $W$ and using the relation  Eq.~(\ref{coupling}) we
evaluated numerically the value of the coupling constant $g$. The obtained
result is in an excellent agreement with the estimated value (\ref{g1}) for all
considered types of the waveguide attachment leading to $g\approx 0.98$. We
have evaluated the coupling constant $g$ also for a different shape of the
billiard \cite{P+98} obtaining similar values for $g$. It has to be stressed
that the estimate (\ref{g1}) was obtained using semiclassical arguments. Our
calculation shows, however, that it remains valid  even in the deep quantum
region.

The eigenvalues of $H_{\rm eff}$ (Eq.~\ref{seffh2}) are usually interpreted as
the resonance poles and are used for the study of the statistical properties
of resonances in open quantum chaotic systems \cite{FS97}. To check the
validity of this approach we have evaluated the resonance poles of the system
independently using the complex scaling method \cite{S+00} that provides a
direct access to the positions of the poles of the analytically continued
$S$-matrix.  The obtained results were compared with the eigenvalues of the
effective Hamiltonian (\ref{seffh2}).  However, a direct comparison is
obscured by the fact that the coupling matrix $W$ is in fact energy dependent.
In the standard random matrix approach the coupling matrix is treated as being
energy independent---a simplification that is well justified inside a small
energy interval.  To mimic this situation and to minimize the influence of the
energy dependence of the coupling matrix $W$ we have compared the eigenvalues
of $H_{\rm eff}$ with the directly evaluated resonance poles always within a
small energy interval the center of which was equal to the energy used to
evaluate the coupling matrix $W$.

\def\figPoles{
 \begin{figure}[htb]
  \centerline{\psfig{figure=\figdir/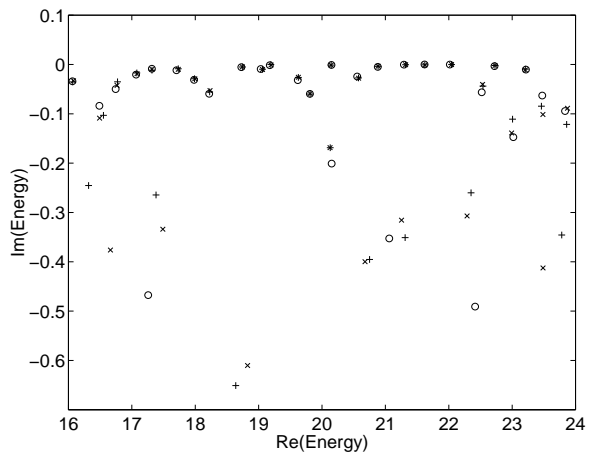,width=\fw}}
  \caption{\label{poles}The complex eigenvalues of $H_{\rm eff}$ (+),
  zeros of $I+\i K^{\rm (N)}$ (x) and resonance poles obtained by the complex
  scaling method (o).}
 \end{figure}
}
\if\tc{\def\fw{78mm}\figPoles}\fi
The energy dependence of $W$ can be taken into account more precisely using
the relation (\ref{sk}) and evaluating the Heidelberg resonance poles as zeros
of the function $I+\i K^{\rm (N)}(E)$. We have evaluated the Heidelberg
resonance poles using both of the above described approaches and compared the
results to the resonances obtained by complex scaling. The results are shown
in Fig.~\ref{poles}. From this figure we see that for narrow resonances the
eigenvalues of effective Hamiltonian $H_{\rm eff}$ represent a good
approximation to the resonance poles of the system and the energy dependence
of the coupling matrix $W$ can be omitted. For broad resonances the situation
changes and the complex eigenvalues of the effective Hamiltonian have nothing
in common with the directly evaluated resonance poles. This discrepancy can be
explained as follows: A resonance localized at $E_R=E+i \Gamma$ represents in
fact a collective mode of all bound states $E_n$ of the internal Hamiltonian
$H_{in}$ that are located inside the energy interval $\approx
(E-\Gamma,E+\Gamma )$. For a broad resonance with $\Gamma$ significantly
larger then the mean spacing between the bound states $E_n$ the number of the
internal states to be included into $H_{\rm eff}$ must be very high. 
\section{Conclusions}
To summarize, we have shown that the Heidelberg approach to scattering, which
is the basis of many important random-matrix results on quantum chaotic
scattering, leads to reasonably good agreement when compared with the results
of a direct calculation of the S-matrix in the case of billiards with Neumann
boundary conditions. We have explained the somewhat unexpected finding that
the accuracy is much worse for Dirichlet boundary conditions, while it does
not depend very much on other possibilities of varying the auxiliary closed
system used, such as the position of the attached waveguide. 

Even for Neumann boundary conditions the effective Hamiltonian $H_{\rm eff}$
based on the Heidelberg approach seems to be not very well suited for the
computation of broad resonances of the system. This result does not contradict
the fact, that also for systems with time-reversal symmetry the {\em
statistical} properties of billiard resonances follow the predictions of
random-matrix theory based upon the effective Hamiltonian approach quite well
\cite{Ish00}, because our test goes way beyond a purely statistical analysis.

Moreover, we have shown numerically and with semiclassical arguments that the
mean dimensionless coupling for a chaotic billiard is the critical value
$g=1$---irrespective of the precise form of the billiard, the size of the
attached waveguide and other details of the model. Interestingly, in our model
an effective coupling near $g=1$ is observed already deep in the quantum
regime. Nevertheless, fluctuations around the mean value $g=1$ should in
general be largest for small energies and can possibly result locally in
overcritical coupling. This might be an explanation for the observed resonance
trapping in billiards \cite{P+98,R+00,S+00,P+00}.

Our results concerning the value of the effective coupling for chaotic systems
are not restricted to billiards and apply, e.~g., to quantum graphs as
well. In these systems, a systematic way to achieve overcritical coupling for
many states is to modulate the density of states, e.~g.\ by considering
systems with band spectra \cite{Dit00,SD01}.\\[5mm]

\begin{acknowledgments}
This work has been partially supported by
the Czech grand GAAV A1048804 and by the "Foundation for
Theoretical Physics" in Slemeno, Czech Republic.
\end{acknowledgments}


\begin{thebibliography}{10}
\expandafter\ifx\csname bibnamefont\endcsname\relax
  \def\bibnamefont#1{#1}\fi
\expandafter\ifx\csname bibfnamefont\endcsname\relax
  \def\bibfnamefont#1{#1}\fi
\expandafter\ifx\csname url\endcsname\relax
  \def\url#1{\texttt{#1}}\fi
\expandafter\ifx\csname urlprefix\endcsname\relax\def\urlprefix{URL }\fi
\providecommand{\bibinfo}[2]{#2}
\providecommand{\eprint}[2][]{\url{#2}}

\bibitem{MW69}
\bibinfo{author}{\bibfnamefont{C.}~\bibnamefont{Mahaux}} \bibnamefont{and}
  \bibinfo{author}{\bibfnamefont{H.~A.} \bibnamefont{Weidenm{\"u}ller}},
  \emph{\bibinfo{title}{Shell-Model Approach to Nuclear Reactions}}
  (\bibinfo{publisher}{North Holland}, \bibinfo{address}{Amsterdam},
  \bibinfo{year}{1969}).

\bibitem{Feshbach}
\bibinfo{author}{\bibfnamefont{H.}~\bibnamefont{Feshbach}},
  \emph{\bibinfo{title}{Theoretical Nuclear Physics}}
  (\bibinfo{publisher}{Wiley-Interscience}, \bibinfo{year}{1992}).

\bibitem{Dit00}
\bibinfo{author}{\bibfnamefont{F.~M.} \bibnamefont{Dittes}},
  \bibinfo{journal}{Phys.~Rep.} \textbf{\bibinfo{volume}{339}},
  \bibinfo{pages}{216} (\bibinfo{year}{2000}).

\bibitem{BGS84}
\bibinfo{author}{\bibfnamefont{O.}~\bibnamefont{Bohigas}},
  \bibinfo{author}{\bibfnamefont{M.~J.} \bibnamefont{Giannoni}},
  \bibnamefont{and} \bibinfo{author}{\bibfnamefont{C.}~\bibnamefont{Schmit}},
  \bibinfo{journal}{Phys. Rev. Lett.} \textbf{\bibinfo{volume}{52}},
  \bibinfo{pages}{1} (\bibinfo{year}{1984}).

\bibitem{MPS85}
\bibinfo{author}{\bibfnamefont{P.~A.} \bibnamefont{Mello}},
  \bibinfo{author}{\bibfnamefont{P.}~\bibnamefont{Pereyra}}, \bibnamefont{and}
  \bibinfo{author}{\bibfnamefont{T.~H.} \bibnamefont{Seligman}},
  \bibinfo{journal}{Ann.~Phys.} \textbf{\bibinfo{volume}{161}},
  \bibinfo{pages}{254} (\bibinfo{year}{1985}).

\bibitem{LW91}
\bibinfo{author}{\bibfnamefont{C.~H.} \bibnamefont{Lewenkopf}}
  \bibnamefont{and} \bibinfo{author}{\bibfnamefont{H.~A.}
  \bibnamefont{Weidenmuller}}, \bibinfo{journal}{Ann.~Phys.}
  \textbf{\bibinfo{volume}{212}}, \bibinfo{pages}{53} (\bibinfo{year}{1991}).

\bibitem{FS97}
\bibinfo{author}{\bibfnamefont{Y.~V.} \bibnamefont{Fyodorov}} \bibnamefont{and}
  \bibinfo{author}{\bibfnamefont{H.~J.} \bibnamefont{Sommers}},
  \bibinfo{journal}{J.~Math.~Phys.} \textbf{\bibinfo{volume}{38}},
  \bibinfo{pages}{1918} (\bibinfo{year}{1997}).

\bibitem{GMW98}
\bibinfo{author}{\bibfnamefont{T.}~\bibnamefont{Guhr}},
  \bibinfo{author}{\bibfnamefont{A.}~\bibnamefont{Muller-Groeling}},
  \bibnamefont{and} \bibinfo{author}{\bibfnamefont{H.~A.}
  \bibnamefont{Weidenmuller}}, \bibinfo{journal}{Phys.~Rep.}
  \textbf{\bibinfo{volume}{299}}, \bibinfo{pages}{190} (\bibinfo{year}{1998}).

\bibitem{SZ89}
\bibinfo{author}{\bibfnamefont{V.~V.} \bibnamefont{Sokolov}} \bibnamefont{and}
  \bibinfo{author}{\bibfnamefont{V.~G.} \bibnamefont{Zelevinsky}},
  \bibinfo{journal}{Nucl.~Phys.~A} \textbf{\bibinfo{volume}{504}},
  \bibinfo{pages}{562} (\bibinfo{year}{1989}).

\bibitem{SZ92}
\bibinfo{author}{\bibfnamefont{V.~V.} \bibnamefont{Sokolov}} \bibnamefont{and}
  \bibinfo{author}{\bibfnamefont{V.~G.} \bibnamefont{Zelevinsky}},
  \bibinfo{journal}{Ann.~Phys.} \textbf{\bibinfo{volume}{216}},
  \bibinfo{pages}{323} (\bibinfo{year}{1992}).

\bibitem{Rot91}
\bibinfo{author}{\bibfnamefont{I.}~\bibnamefont{Rotter}},
  \bibinfo{journal}{Rep.~Prog.~Phys.} \textbf{\bibinfo{volume}{54}},
  \bibinfo{pages}{635} (\bibinfo{year}{1991}).

\bibitem{DS92a}
\bibinfo{author}{\bibfnamefont{E.}~\bibnamefont{Doron}} \bibnamefont{and}
  \bibinfo{author}{\bibfnamefont{U.}~\bibnamefont{Smilansky}},
  \bibinfo{journal}{Phys. Rev. Lett.} \textbf{\bibinfo{volume}{68}},
  \bibinfo{pages}{1255} (\bibinfo{year}{1992}).

\bibitem{M+92}
\bibinfo{author}{\bibfnamefont{C.~M.} \bibnamefont{Marcus}},
  \bibinfo{author}{\bibfnamefont{A.~J.} \bibnamefont{Rimberg}},
  \bibinfo{author}{\bibfnamefont{R.~M.} \bibnamefont{Westervelt}},
  \bibinfo{author}{\bibfnamefont{P.~F.} \bibnamefont{Hopkins}},
  \bibnamefont{and} \bibinfo{author}{\bibfnamefont{A.~C.}
  \bibnamefont{Gossard}}, \bibinfo{journal}{Phys.~Rev.~Lett.}
  \textbf{\bibinfo{volume}{69}}, \bibinfo{pages}{506} (\bibinfo{year}{1992}).

\bibitem{SS90}
\bibinfo{author}{\bibfnamefont{H.-J.} \bibnamefont{St\"ockmann}}
  \bibnamefont{and} \bibinfo{author}{\bibfnamefont{J.}~\bibnamefont{Stein}},
  \bibinfo{journal}{Phys. Rev. Lett.} \textbf{\bibinfo{volume}{64}},
  \bibinfo{pages}{2215} (\bibinfo{year}{1990}).

\bibitem{P+00}
\bibinfo{author}{\bibfnamefont{E.}~\bibnamefont{Persson}},
  \bibinfo{author}{\bibfnamefont{I.}~\bibnamefont{Rotter}},
  \bibinfo{author}{\bibfnamefont{H.~J.} \bibnamefont{St{\"o}ckmann}},
  \bibnamefont{and} \bibinfo{author}{\bibfnamefont{M.}~\bibnamefont{Barth}},
  \bibinfo{journal}{Phys. Rev. Lett.} \textbf{\bibinfo{volume}{85}},
  \bibinfo{pages}{2478} (\bibinfo{year}{2000}).

\bibitem{P+98}
\bibinfo{author}{\bibfnamefont{E.}~\bibnamefont{Persson}},
  \bibinfo{author}{\bibfnamefont{K.}~\bibnamefont{Pichugin}},
  \bibinfo{author}{\bibfnamefont{I.}~\bibnamefont{Rotter}}, \bibnamefont{and}
  \bibinfo{author}{\bibfnamefont{P.}~\bibnamefont{Seba}},
  \bibinfo{journal}{Phys.~Rev.~E} \textbf{\bibinfo{volume}{58}},
  \bibinfo{pages}{8001} (\bibinfo{year}{1998}).

\bibitem{R+00}
\bibinfo{author}{\bibfnamefont{I.}~\bibnamefont{Rotter}},
  \bibinfo{author}{\bibfnamefont{E.}~\bibnamefont{Persson}},
  \bibinfo{author}{\bibfnamefont{K.}~\bibnamefont{Pichugin}}, \bibnamefont{and}
  \bibinfo{author}{\bibfnamefont{P.}~\bibnamefont{Seba}},
  \bibinfo{journal}{Phys.~Rev.~E} \textbf{\bibinfo{volume}{62}},
  \bibinfo{pages}{450} (\bibinfo{year}{2000}).

\bibitem{S+00}
\bibinfo{author}{\bibfnamefont{P.}~\bibnamefont{Seba}},
  \bibinfo{author}{\bibfnamefont{I.}~\bibnamefont{Rotter}},
  \bibinfo{author}{\bibfnamefont{M.}~\bibnamefont{Muller}},
  \bibinfo{author}{\bibfnamefont{E.}~\bibnamefont{Persson}}, \bibnamefont{and}
  \bibinfo{author}{\bibfnamefont{K.}~\bibnamefont{Pichugin}},
  \bibinfo{journal}{Phys.~Rev.~E} \textbf{\bibinfo{volume}{61}},
  \bibinfo{pages}{66} (\bibinfo{year}{2000}).

\bibitem{DS92b}
\bibinfo{author}{\bibfnamefont{E.}~\bibnamefont{Doron}} \bibnamefont{and}
  \bibinfo{author}{\bibfnamefont{U.}~\bibnamefont{Smilansky}},
  \bibinfo{journal}{Nonlinearity} \textbf{\bibinfo{volume}{5}},
  \bibinfo{pages}{1055} (\bibinfo{year}{1992}).

\bibitem{SS95}
\bibinfo{author}{\bibfnamefont{H.}~\bibnamefont{Schanz}} \bibnamefont{and}
  \bibinfo{author}{\bibfnamefont{U.}~\bibnamefont{Smilansky}},
  \bibinfo{journal}{Chaos, Soli\-tons $\&$ Fractals}
  \textbf{\bibinfo{volume}{5}}, \bibinfo{pages}{1289} (\bibinfo{year}{1995}).

\bibitem{BH76}
\bibinfo{author}{\bibfnamefont{H.~P.} \bibnamefont{Baltes}} \bibnamefont{and}
  \bibinfo{author}{\bibfnamefont{E.~R.} \bibnamefont{Hilf}},
  \emph{\bibinfo{title}{Spectra of Finite Systems}}
  (\bibinfo{publisher}{Bibliographisches Institut},
  \bibinfo{address}{Mannheim}, \bibinfo{year}{1976}).

\bibitem{A91}
\bibinfo{author}{\bibfnamefont{T.}~\bibnamefont{Ando}},
  \bibinfo{journal}{Phys.~Rev.~B} \textbf{\bibinfo{volume}{44}},
  \bibinfo{pages}{8017} (\bibinfo{year}{1991}).

\bibitem{Ish00}
\bibinfo{author}{\bibfnamefont{H.}~\bibnamefont{Ishio}},
  \bibinfo{journal}{Phys.~Rev.~E} \textbf{\bibinfo{volume}{62}},
  \bibinfo{pages}{R3035} (\bibinfo{year}{2000}).

\bibitem{SD01}
\bibinfo{author}{\bibfnamefont{H.}~\bibnamefont{Schanz}} \bibnamefont{and}
  \bibinfo{author}{\bibfnamefont{F.~M.} \bibnamefont{Dittes}},
  \bibinfo{note}{to be published}.

\end{thebibliography}


\if\tc\else
\def\fw{150mm}\figSketch\newpage
\def\fw{150mm}\figGeom\newpage
\def\fw{150mm}\figTDlow\newpage
\def\fw{150mm}\figPoles
\fi
\end{document}